\documentclass[amsmath,amssymb,preprint,aps]{revtex4}
\renewcommand{\baselinestretch}{1.2}
\input psfig
\begin{document}
\title{Three-Neutrino Mixing after the First Results from K2K and KamLAND}
\author{M. C. Gonzalez-Garcia
\footnote{E-mail: concha@insti.physics.sunysb.edu}}
\affiliation{ Y.I.T.P., SUNY at Stony Brook, Stony Brook, NY 11794-3840, USA\\
IFIC, Universitat de Val\`encia - C.S.I.C., Apt 22085, 46071
Val\`encia, Spain}
\author{Carlos Pe\~na-Garay\footnote{E-mail: penya@ias.edu}}
\affiliation{School of Natural Sciences, Institute for Advanced Study,
Princeton, NJ 08540, USA
\vspace*{2cm}}
\begin{abstract}
\vspace*{1cm}
We analyze the impact of the data 
on long baseline $\nu_\mu$  disappearance from the K2K experiment 
and reactor $\bar\nu_e$ disappearance from the KamLAND experiment
on the determination of the leptonic three-generation mixing 
parameters. Performing an up-to-date global analysis  of
solar, atmospheric, reactor and long baseline neutrino  
data in the context of three-neutrino oscillations, we determine 
the presently allowed ranges of masses and mixing and we consistently derive 
the allowed magnitude of the elements of the leptonic mixing matrix. 
We also quantify the maximum allowed contribution of $\Delta m^2_{21}$
oscillations to CP-odd and CP-even observables at future long 
baseline experiments.
\end{abstract}
\maketitle

\renewcommand{\baselinestretch}{1.2}
\section{Introduction}
Neutrino oscillations are entering in a new era in which
the observations from underground experiments obtained with
neutrino beams provided to us by Nature -- either from the
Sun or from the interactions of cosmic rays in the upper
atmosphere-- are being confirmed by experiments using 
``man-made'' neutrinos from accelerators and nuclear reactors. 

Super--Kamiokande (SK) high statistics data~\cite{skatmlast,skatmpub} 
clearly established that the observed deficit in the $\mu$-like atmospheric 
events is due to the neutrinos arriving in the detector at large 
zenith angles, strongly suggestive of the $\nu_\mu$ oscillation 
hypothesis. This evidence was also confirmed by other atmospheric
experiments such as MACRO~\cite{macro} and Soudan 2~\cite{soudan}.
Similarly, the SNO results~\cite{snoccnc} in combination  with the 
SK data on the zenith angle dependence and recoil energy
spectrum of solar neutrinos~\cite{sksollast} and the 
Homestake~\cite{chlorine}, SAGE~\cite{sage}, 
GALLEX+GNO~\cite{gallex,gno} and Kamiokande~\cite{kamiokande} 
experiments, put on a firm observational basis the long--standing problem 
of solar neutrinos~\cite{snp}, establishing the need for $\nu_e$
conversions. 

The KEK to Kamioka long-baseline neutrino oscillation experiment
(K2K) uses an accelerator-produced neutrino beam mostly consisting
of $\nu_\mu$
with a mean energy of 1.3~GeV and a neutrino flight distance 
of 250~km to probe the same 
oscillations that were explored with atmospheric neutrinos.  Their 
results~\cite{k2kprl} show that
both the number of observed neutrino events and the observed energy
spectrum are consistent with neutrino oscillations with 
oscillation parameters consistent with the ones suggested by 
atmospheric neutrinos.

The KamLAND experiment measures the flux of $\bar{\nu}_e$'s from
nuclear reactors with a energy of $\sim$ MeV located at typical 
distance of  $\sim$ 180~km  with the aim of exploring with a terrestrial 
beam the region of neutrino parameters that is relevant for 
the oscillation interpretation of the solar data.  
Their first published~\cite{kamland} results show that
both the total number of events and their energy spectrum 
can be better interpreted in terms of $\bar\nu_e$  oscillations 
with parameters consistent with the the LMA solar neutrino solution
~\cite{kamland,oursolar,ourkland,othersolkam}. 

Altogether, the data from solar and atmospheric neutrino
experiments, and the first results from KamLAND and
K2K constitute the only solid present--day evidence for physics 
beyond the Standard Model \cite{review}. The minimum joint 
description of these data requires neutrino mixing among 
all three known neutrinos and it determines the structure of the 
lepton mixing matrix~\cite{MNS} which can be parametrized 
as~\cite{PDG}
\begin{equation}
U=\left(
    \begin{array}{ccc}
        c_{13} c_{12}
        & s_{12} c_{13}
        & s_{13} \,{e}^{-i\delta}\\
        -s_{12} c_{23} - s_{23} s_{13} c_{12} \,{e}^{i\delta}
        & c_{23} c_{12} - s_{23} s_{13} s_{12}\, {e}^{i\delta}
        & s_{23} c_{13} \\
        s_{23} s_{12} - s_{13} c_{23} c_{12} \,{e}^{i\delta}
        & -s_{23} c_{12} - s_{13} s_{12} c_{23} \, {e}^{i\delta}
        & c_{23} c_{13}
    \end{array} \right) \,,
    \label{eq:matrix}
\end{equation}
where $c_{ij} \equiv \cos\theta_{ij}$ and $s_{ij} \equiv
\sin\theta_{ij}$. 
In addition to the Dirac-type phase, $\delta$, analogous to that of the
quark sector, there are two physical phases associated to the Majorana
character of neutrinos, which however are not relevant for neutrino
oscillations \cite{majophases} and will be set to zero in what follows. 
 
In this paper we present the result of the global analysis  of
solar, atmospheric, reactor and long-baseline neutrino  
data in the context of three-neutrino oscillations with the aim of 
determining in a consistent way our present knowledge 
of the leptonic mixing matrix  and the neutrino
mass differences. We make particular emphasis on 
the impact of the first data from long baseline $\nu_\mu$  disappearance 
from the K2K experiment and reactor $\bar\nu_e$ disappearance from 
the KamLAND experiment. 

The outline of the paper is as follows. In Sec.~\ref{data} we
describe the data included in the analysis and we briefly describe
the relevant formalism. Sec.~\ref{k2k} contains the results 
of the analysis of the 
K2K data and their effect on the determination of the
parameters associated with atmospheric  oscillations. 
We find that the main impact of K2K  when combined with 
atmospheric neutrino data is to reduce the allowed
range of the corresponding mass difference. When combined with the data 
from the CHOOZ~\cite{chooz} 
experiment in a three-neutrino analysis, this
results into a slight tightening of the derived bound on 
$\theta_{13}$ at high CL. In Sec.~\ref{osc12} 
we describe the results from the global analysis including also 
solar and KamLAND data and  in Sec.~\ref{matrix}  we describe our 
procedure to consistently derive 
the allowed magnitude of the elements of the leptonic mixing matrix.
As outcome of this analysis we also  quantify the maximum allowed 
contribution 
of $\Delta m^2_{21}$ oscillations to CP-odd and CP-even observables 
at future long  baseline experiments in Sec.~\ref{lbl} .
Conclusions are given in Sec.~\ref{conclu}. 
We also present an appendix with the
details of our analysis of the K2K data. 

\section{Data Inputs and Formalism}
\label{data}
We include in our statistical analysis the data from solar, atmospheric
and K2K accelerator  neutrinos and from the CHOOZ and KamLAND reactor 
antineutrinos. 

In the analysis of K2K we include the data on the normalization and 
shape of the spectrum  of single-ring $\mu$-like events as a function 
of the reconstructed neutrino energy. The total sample corresponds 
to 29 events.  In the absence of oscillations 44 events were expected. We bin
the data  in five 0.5 GeV bins with $0<E_{\rm rec}<2.5$ plus 
one bin containing all events above $2.5$ GeV. 
For QE events the reconstructed neutrino energy is well distributed
around the true neutrino energy, However experimental energy and 
angular resolution and more importantly the contamination from 
non-QE events, 
result into important deviations of the reconstructed neutrino energy
from the true neutrino energy which we carefully account for. 
We include the systematic uncertainties associated with the determination 
of the neutrino energy  spectrum in the near detector, 
the model dependence of the amount of nQE contamination, 
the  near/far extrapolation  and the overall flux normalization.
Details of this analysis are presented in the appendix.

For atmospheric neutrinos we include in our analysis all the
contained events from the latest 1489 SK data
set~\cite{skatmlast}, as well as the upward-going neutrino-induced
muon fluxes from both SK and the MACRO detector~\cite{macro}.
This amounts for a total of 65 data points. 
More technical description of our simulations and statistical analysis 
can be found in Refs.~\cite{ouratmos,3ours}.

We refine our previous analysis~\cite{3ours,nohier} of the 
CHOOZ reactor data~\cite{chooz}  and include
here their energy binned data instead of their total rate only. 
This corresponds  to 14 data points (7-bin positron spectra from
both reactors, Table 4 in Ref.~\cite{chooz}) with one constrained 
normalization parameter and including all the systematic uncertainties
there described. 

For the solar neutrino analysis, we use 80 data points.  We include
the two measured radiochemical rates, from the
chlorine~\cite{chlorine} and the gallium~\cite{sage,gallex,gno}
experiments, the 44 zenith-spectral energy bins of the electron
neutrino scattering signal measured by the SK
collaboration ~\cite{sksollast}, and the 34 day-night spectral energy
bins measured with the SNO~\cite{snoccnc}
detector. 
We take account of the BP00~\cite{bp00} predicted fluxes and
uncertainties for all solar neutrino sources except for $^8$B
neutrinos. We treat the total
$^8$B solar neutrino flux as a free parameter to be determined by
experiment and to be compared with solar model predictions.
For KamLAND we include information on the observed antineutrino 
spectrum which accounts for a total of 13 data points. 
Details of our calculations and 
statistical treatment of solar and KamLAND data can be found in 
Refs.~\cite{oursolar,ourkland}.

In general the parameter set relevant for the joint study
of these neutrino data in the framework of three-$\nu$ mixing 
is six-dimensional: two mass differences, three mixing angles 
and one CP phase. 

Results from the analysis of solar plus KamLAND, and atmospheric data 
in the framework of oscillations between two neutrino states~\cite{oursolar,
ourkland,othersolkam,ouratmos,otheratmos} imply that the required 
mass differences satisfy that 
\begin{equation}
\Delta m^2_\odot\ll \Delta m^2_{\rm atm}.
\label{deltahier}
\end{equation}
In this approximation the angles $\theta_{ij}$ in Eq.~(\ref{eq:matrix})
can be taken without 
loss of generality to lie in the first quadrant, $\theta_{ij}\in[0,\pi/2]$.  
There are two possible mass orderings which we chose as
\begin{eqnarray}
\Delta m^2_{21}=\Delta m^2_\odot &\ll&
\Delta m^2_{32}\simeq\Delta m^2_{31}=\Delta m^2_{\rm atm}>0; \label{direct}\\
\Delta m^2_{21}=\Delta m^2_\odot &\ll&
-\Delta m^2_{31}\simeq-\Delta m^2_{32}=|\Delta m^2_{\rm atm}|>0. 
\label{inverted}
\end{eqnarray}
As it is customary we refer to the first option, Eq.~(\ref{direct}), as the 
{\it normal} scheme,
and to the second one, Eq.~(\ref{inverted}), as the {\it inverted} scheme.  

For solar neutrinos and for antineutrinos in KamLAND 
the oscillations with $\Delta m^2_{32}\sim\Delta m^2_{31}$ 
are averaged out. The relevant survival probability takes 
the form:
\begin{equation}
P^{3\nu}_{ee}
=\sin^4\theta_{13}+ \cos^4\theta_{13}P^{2\nu}_{ee} 
(\Delta m^2_{21},\theta_{12}) \; , 
\label{pee3}
\end{equation}
where 
$P^{2\nu}_{ee} (\Delta m^2_{21},\theta_{12})$ is the survival probability
for 2$\nu$ mixing, which, for solar neutrinos, is obtained with the modified 
sun density $N_{e}\rightarrow \cos^2\theta_{13} N_e$.
So the analysis of solar and KamLAND data depends on 
three of the five oscillation parameters: $\Delta
m^2_{21}, \theta_{12}$ and $\theta_{13}$. 

Conversely for small $\Delta m^2_{21}$ the three-neutrino
oscillation analysis of the atmospheric and K2K neutrino data can be performed
in the one mass scale dominance approximation neglecting the
effect of $\Delta m^2_{21}$.
In this approximation the angle $\theta_{12}$ can be rotated away and 
it follows that the atmospheric and K2K data analysis restricts three 
of the oscillation
parameters, namely, 
$\Delta m^2_{31} = \Delta m^2_{32}$, $\theta_{23}$
and $\theta_{13}$ and the CP phase $\delta$ becomes unobservable.
The $\nu_\mu$ survival probability at K2K is 
\begin{eqnarray}
P^{\rm K2K}_{\mu\mu}&=& 1-4 \left(s^4_{23} s^2_{13} c^2_{13}
          + c^2_{13} s^2_{23} c^2_{23}\right) 
\sin^2\left(\frac{\Delta m^2_{32}}{4E_\nu}L\right) \nonumber\\
&\simeq& s_{13}^2\frac{\cos2\theta_{23}}{c_{23}^2}
+(1-s_{13}^2\frac{\cos2\theta_{23}}{c_{23}^2})
P^{K2K,2\nu}_{\mu\mu}(\Delta m^2_{32},\theta_{32})\; +\;{\cal O}(s_{13}^4)\; .
\label{eq:pk2k}
\end{eqnarray}

For atmospheric neutrinos in the general case of three-neutrino scenario with
$\theta_{13}\neq 0$  the presence of the matter
potentials become relevant. We solve numerically  the
evolution equations in order to obtain the oscillation probabilities
for both e and $\mu$ flavours, which are different for neutrinos 
and anti-neutrinos. Because of the matter effects, they also
depend on the mass ordering being normal or inverted.
In our calculations, we use for the matter
density profile of the Earth the approximate analytic parametrization
given in Ref.~\cite{lisi} of the PREM model of the Earth~\cite{PREM}.

The reactor neutrino data from CHOOZ provides information on the 
survival probability~\cite{3ours,choozpetcov,irina}.  
\begin{eqnarray}
P_{ee}^{\rm CHOOZ}&=& 1-c^4_{13}\sin^22\theta_{12}
\sin^2\left(\frac{\Delta m^2_{21} L}{4 E_\nu} \right) 
-\sin^22\theta_{13} 
\left[c^2_{12}\sin^2\left(\frac
{\Delta m^2_{31} L}{4 E_\nu}\right) \right.
\nonumber \\
&&\left.
+s^2_{12}\sin^2\left(\frac
{\Delta m^2_{32} L}{4 E_\nu}\right)\right]
\simeq 1-\sin^22\theta_{13}\sin^2\left(\frac{\Delta m^2_{32}L}{4E_\nu}\right).
\label{pchooz} 
\end{eqnarray}
The second equality holds in the approximation
$\Delta m^2_{21}\ll E_\nu/L$ which can be safely made 
for the presently allowed values of $\Delta m^2_{21}$
~\cite{ourkland,othersolkam}. Thus  
the analysis of the CHOOZ reactor data involves only two parameters: 
$\Delta m^2_{32}$ and the mixing angle $\theta_{13}$. 

In summary, oscillations at solar+KamLAND on one side, and atmospheric+K2K 
oscillations on the other side, 
decouple in the limit $\theta_{13}=0$. In this case the values of 
allowed parameters can be obtained directly
from the results of the analysis in terms of two--neutrino oscillations
and normal and inverted hierarchies are equivalent. 
Deviations from the two--neutrino scenario are determined by the
size of the mixing $\theta_{13}$. 

The allowed ranges of masses and mixing obtained in our two-neutrino 
oscillation analysis of solar+KamLAND data can be found in 
Ref.~\cite{ourkland} and we do not reproduce them here. 
We discuss next the results of our analysis of K2K data and its impact
on the determination of the parameters relevant in atmospheric 
oscillations.

\section{Results}
\subsection{$\Delta m_{32}^2$ Oscillations: Impact of K2K Data}
\label{k2k}

\begin{figure}[ht]
\centerline{\psfig{figure=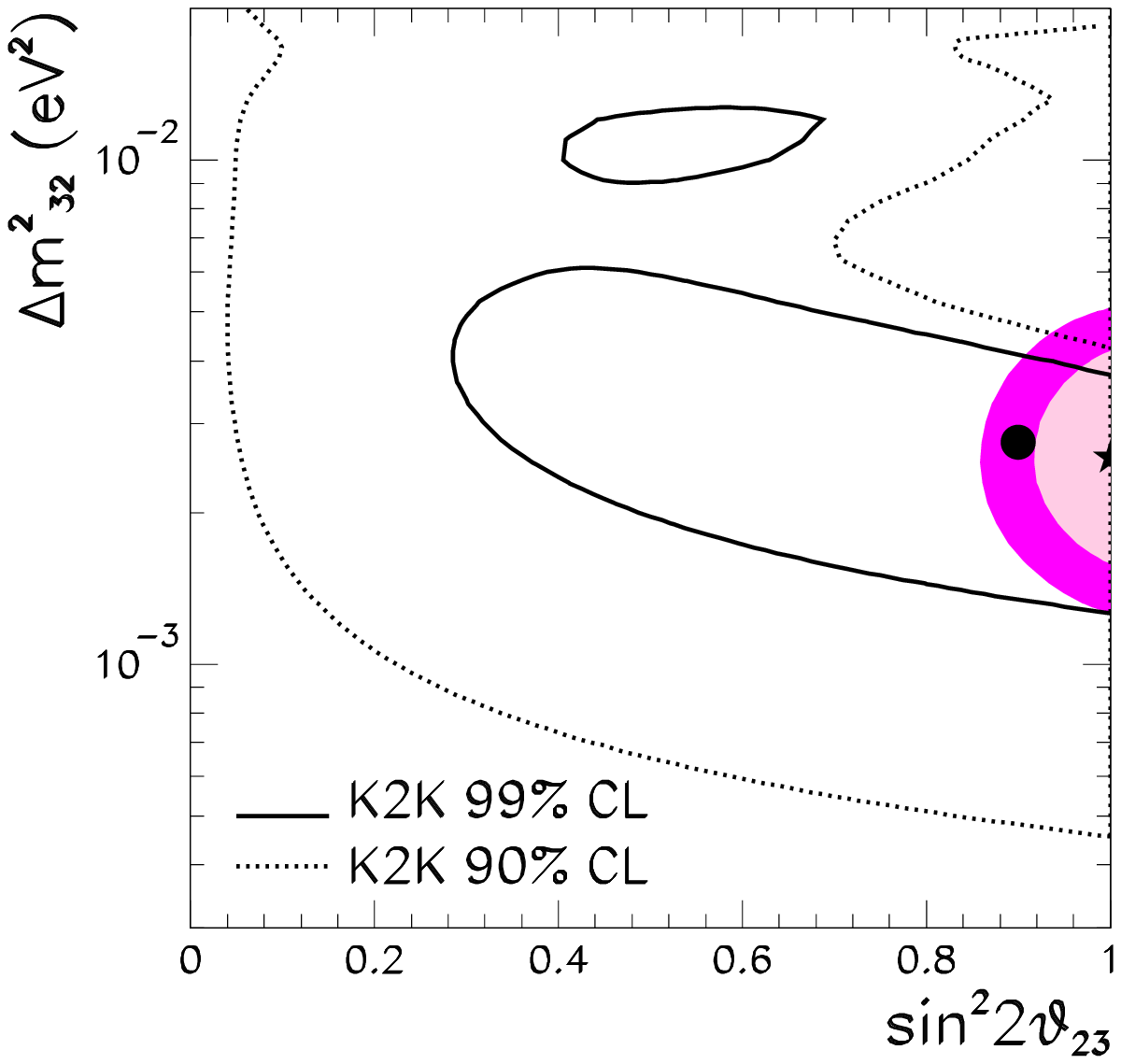,width=3in}
\psfig{figure=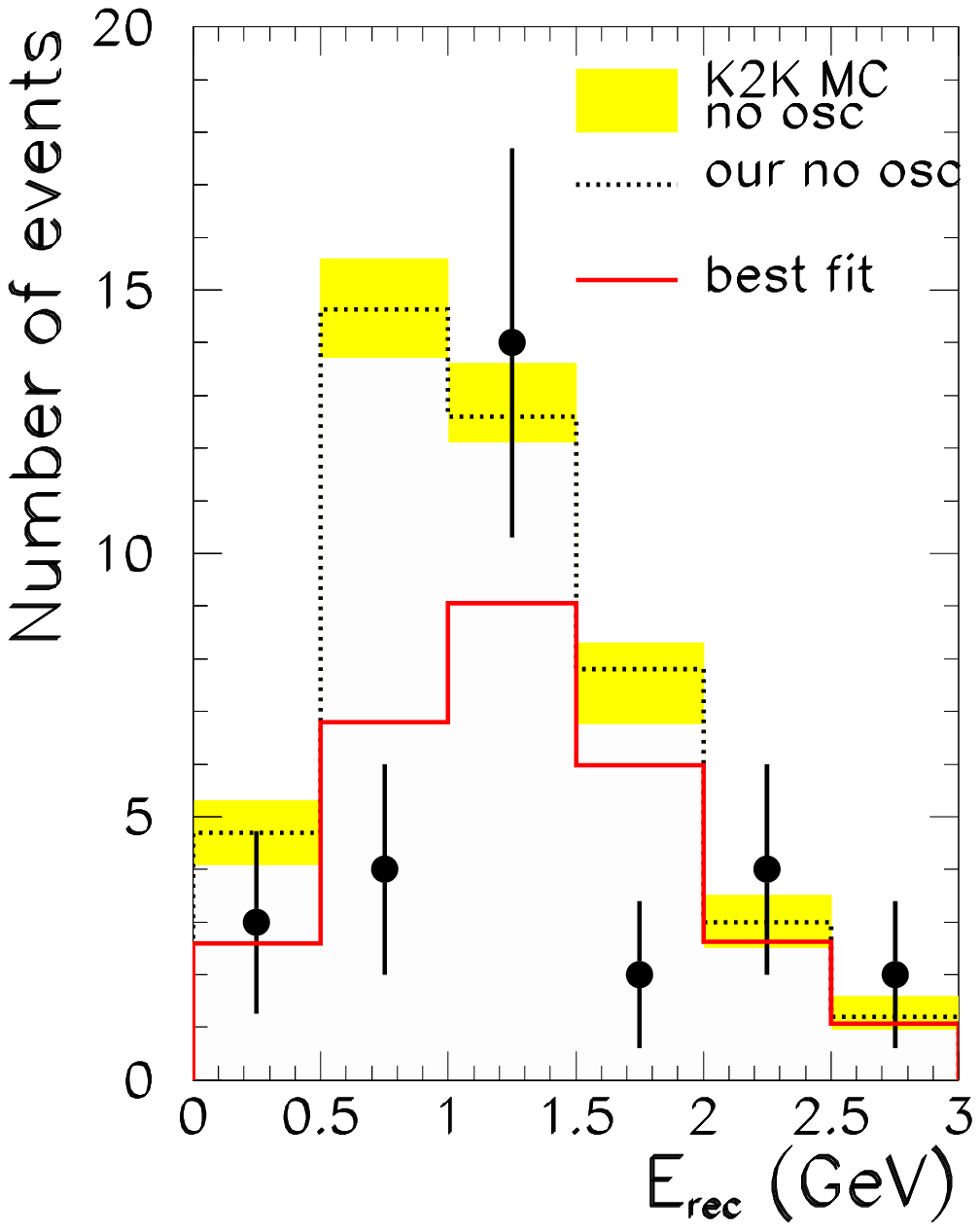,width=2.3in}}
\caption{$\nu_\mu\rightarrow\nu_\tau$ oscillation
analysis of K2K data. On the left panel we show the allowed (2d.o.f) 
regions on $\Delta m^2,\sin^22\theta$ at 90\% CL (solid) and 99\%CL
(dotted). The best fit point is marked with a thick dot.
The shadowed regions are the 90, and 99\% CL of the atmospheric neutrino
analysis with best point marked by a star. 
The right panel shows the spectrum of K2K events as a function of 
the reconstructed neutrino energy for the 6 bins used in the analysis. 
The data points are 
shown together with their statistical errors. The dotted histogram
and the shadow boxes represent our prediction and the K2K MC 
prediction in the absence of oscillations respectively. The full line
represents the expected distribution for the best fit point 
for $\nu_\mu\rightarrow\nu_\tau$ oscillations.} 
\label{fig:k2kdist}
\end{figure}

For sake of comparison with the K2K oscillation analysis, we discuss 
first the results of our analysis of K2K 
data for pure $\nu_\mu \rightarrow\nu_\tau$ oscillations which are
graphically displayed in  Fig.~\ref{fig:k2kdist}. 

We show in the left panel of Fig.~\ref{fig:k2kdist} the allowed region of
$(\Delta m^2, \sin^2(2\theta))$ from our analysis of K2K data.
The  best fit point for this analysis is at 
$\Delta m^2=2.7\times 10^{-3} {\rm eV}^2, 
\sin^2(2\theta)=0.92$ with 
$\chi^2_{min}=9.3$ (the corresponding best fit as obtained from the
K2K collaboration is a $\Delta m^2=2.7\times 10^{-3}{\rm eV}^2 , 
\sin^2(2\theta)=1$). 
We notice  that  the non-maximality of the mixing angle in our analysis 
is not statistically significant as maximal mixing is only at
$\Delta\chi^2=0.15$. The energy spectrum for this point is shown
in the right panel together with the data points and the expectations
in the absence of oscillations. Our results show very good agreement 
with those obtained by the K2K collaboration~\cite{k2kprl}. 
Also displayed in the figure are the corresponding regions from 
our latest atmospheric neutrino analysis~\cite{nohier}. 
As seen in the figure, the K2K results confirm the presence of   
$\nu_\mu$ neutrino oscillations with 
oscillation parameters  consistent with the ones obtained from  
atmospheric neutrinos studies. Furthermore, already at this
first stage, it provides
a restriction on the allowed range of $\Delta m^2$,  
while their dependence on the mixing angle is considerably  weaker. 

In the framework of 3$\nu$ mixing the analysis of K2K, atmospheric
and CHOOZ data provides information on the parameters 
$\Delta m^2_{31}$, $\theta_{23}$, and $\theta_{13}$. We define:
\begin{eqnarray}
\chi^2_{\rm ATM+CHOOZ+K2K}(\Delta m^2_{32},\theta_{23},\theta_{13})
= \chi^2_{\rm ATM} (\Delta m^2_{32},\theta_{23},\theta_{13})
\nonumber\\
+\chi^2_{\rm CHOOZ}(\Delta m^2_{32},\theta_{13}) 
+\chi^2_{\rm K2K} (\Delta m^2_{32},\theta_{23},\theta_{13}) \; .
\label{eq:chi232}
\end{eqnarray}
\begin{figure}[!t]
\centerline{\psfig{figure=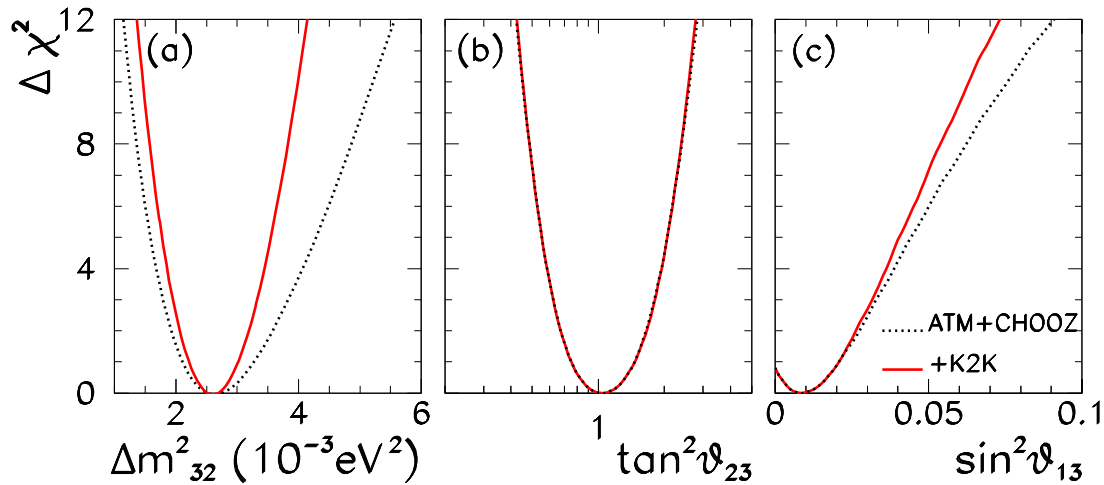,width=6in}} 
\caption{3$\nu$ oscillation analysis of the atmospheric, CHOOZ and
K2K data. The left, center and right panels show the dependence
of $\Delta\chi^2$ on 
$\Delta m^2_{32}$, $\tan^2\theta_{23}$ and $\sin^2\theta_{13}$ 
for the analysis of atmospheric, CHOOZ and K2K data (full line) 
compared to the previous bound before the inclusion of K2K (dotted line).
The individual 1\; (3) $\sigma$ bounds in Eq.~(\ref{akcranges}) 
can be read from the figure with the condition $\Delta\chi^2\leq 1\; (9)$. 
\label{fig:k2k3fam}}
\end{figure}
In the three panels of Fig.~\ref{fig:k2k3fam} we show the bounds on each 
of the three parameters obtained from this analysis (full lines). 
For comparison we also show the corresponding ranges for the analysis
of atmospheric and CHOOZ data alone (dotted lines).
The corresponding subtracted minima  are given in Table~\ref{tab:k2kbest}. 
The results in the figure are shown for the normal mass ordering, but 
once the constraint on $\theta_{13}$ from CHOOZ is included 
in the analysis, the differences between the
results for normal and inverted mass ordering are minimal.  
The careful reader may notice that the $\chi^2$ per d.o.f. 
seems {\sl too good}.
As seen in Table~\ref{tab:k2kbest} this effect is driven by the
atmospheric data and it was already the case for the previous SK 
data sample. It is partly due to the very good agreement of the multi-GeV 
electron distributions with their no-oscillation expectations. However,
as discussed in Ref.~\cite{lisik2k}, $\chi^2_{\rm min}$ is only 2$\sigma$
below its characteristic value, not low enough to be statistically 
suspected.
\begin{table}
\begin{tabular}{|c|c c c|}
\hline
& ATMOS & ATMOS+CHOOZ & ATMOS+CHOOZ+K2K\\
\hline 
Data points & 65 & 65+14=79  &  65+14+6=85 \\
$\frac{\Delta m^2_{32}}{\rm eV^2}$ &  $2.7\times 10^{-3}$ & 
$2.55\times 10^{-3}$ & $2.6\times 10^{-3}$   \\ 
$\tan^2\theta_{23}$  & 1      & 1   &   1 \\
$\sin^2\theta_{13}$  & 0.015  & 0.009 &  0.009    \\
$\chi^2_{\rm min}$   & 39.7   & 45.8     & 55.1 \\
\hline
\end{tabular}
\caption{Minimum $\chi^2$ values and best-fit points for
the $3\nu$ oscillation analysis of atmospheric, CHOOZ and K2K data.}
\label{tab:k2kbest}
\end{table}

In each panel the displayed $\chi^2$ has been marginalized with respect
to the other two parameters.  From the figure we see that the inclusion 
of K2K data in the analysis results into a reduction of the allowed
range of $\Delta m^2_{32}$ while the allowed range of $\theta_{23}$
is not modified. The reduction is more significant for the upper 
bound of $\Delta m^2_{32}$ while the lower bound is slightly increased.
More quantitatively we find that the 
following ranges of  parameters are allowed at 1$\sigma$ (3$\sigma$) 
CL from this analysis
\begin{eqnarray}
(1.5)\; 2.2 \; <\; \Delta m^2_{32}/10^{-3}\mbox{\rm eV$^2$}\;< \; 
\;3.0\, (3.9) \; ,\nonumber \\ 
(0.45)\, 0.75\;<\;\tan^2\theta_{23}\;<\; 1.3 \, (2.3)\; . 
\label{akcranges}
\end{eqnarray}
These ranges are consistent with the results from the 2-neutrino 
oscillation analysis of  K2K and atmospheric data in Ref.~\cite{lisik2k}.

Concerning the ``generic'' 3$\nu$ mixing parameter $\theta_{13}$,
Eq.~(\ref{eq:pk2k}) shows that its effect on both the normalization 
and the shape of the $\nu_\mu$ spectrum is further suppressed 
near maximal mixing by $\cos2\theta_{23}\sim 0$. As a consequence
K2K  alone does not provide any bound on $\theta_{13}$.  
However Fig.~\ref{fig:k2k3fam}.c illustrates how the inclusion
of the long-baseline data results into a tightening of the 
bound on $\theta_{13}$ (at large CL) when combined with the atmospheric and 
CHOOZ data.
This is an indirect effect due to the increase in the lower bound
on $\Delta m^2_{32}$ .
In the favoured range of  $\Delta m^2_{32}$ the oscillating
phase at CHOOZ is small enough so that it can be expanded 
and the oscillation probability of $\bar\nu_e$ depends 
quadratically on  $\Delta m^2_{32}$. As a consequence 
the bound on the mixing angle from CHOOZ is a very sensitive function
of the allowed values for  $\Delta m^2_{32}$.  The increase of the lower
bound on $\Delta m^2_{32}$ due to the inclusion of the K2K data 
leads to the tightening of the derived limit on $\theta_{13}$ at high CL. 
From Fig.~\ref{fig:k2k3fam}.c and Table~\ref{tab:k2kbest} we 
also see that the best fit point is not exactly at  $\sin^2\theta_{13}=0$,
although this is not very statistically significant. This effect is
due to the atmospheric neutrino data. In particular to the slight excess 
of sub GeV e-like events which is better described with a non-vanishing
value of $\theta_{13}$.

\subsection{Global Analysis}
\label{osc12}
\begin{figure}[!t]
\centerline{\psfig{figure=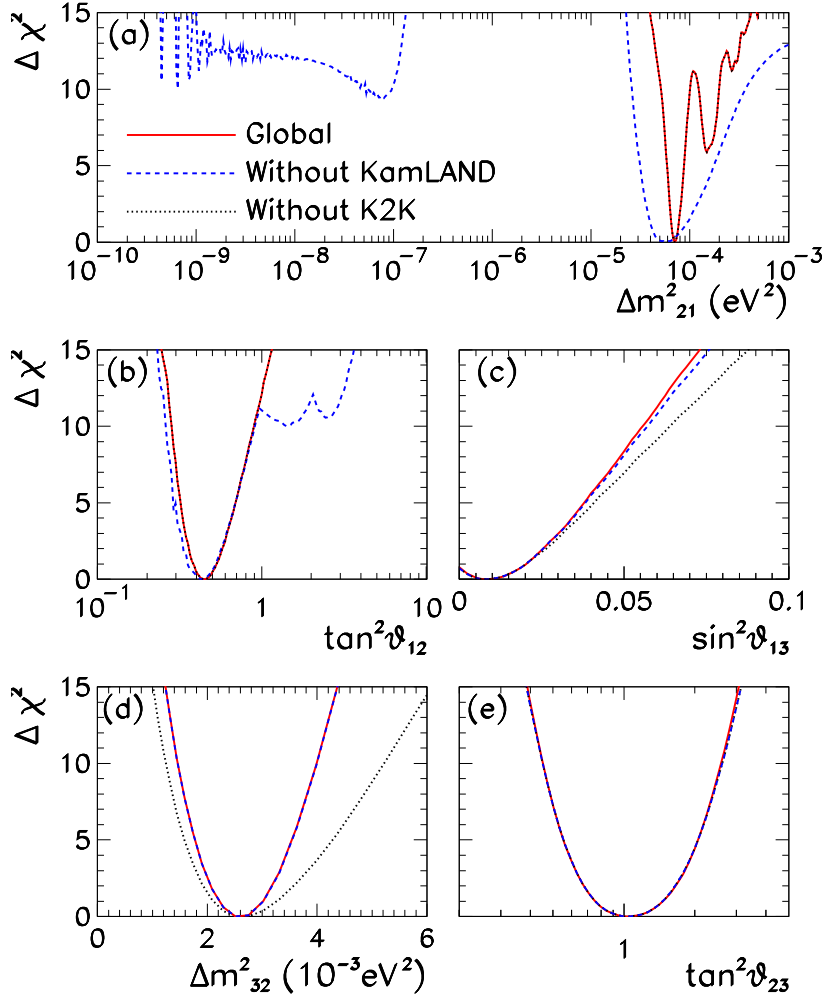,width=4.5in}} 
\caption{Global 3$\nu$ oscillation analysis. Each panels shows the 
dependence of $\Delta\chi^2$ on each of the five parameters from 
the global analysis (full line) compared to the bound prior to the 
inclusion of the K2K (dotted line) and KamLAND data (dashed line).
The individual 1\; (3) $\sigma$ bounds in Eqs.
(\ref{skranges}) and~(\ref{13ranges}) ~(\ref{akcranges}) can be read from 
the corresponding panel with the condition $\Delta\chi^2\leq 1\; (9)$. } 
\label{fig:chiglo}
\end{figure}

\begin{figure}[!t]
\centerline{\psfig{figure=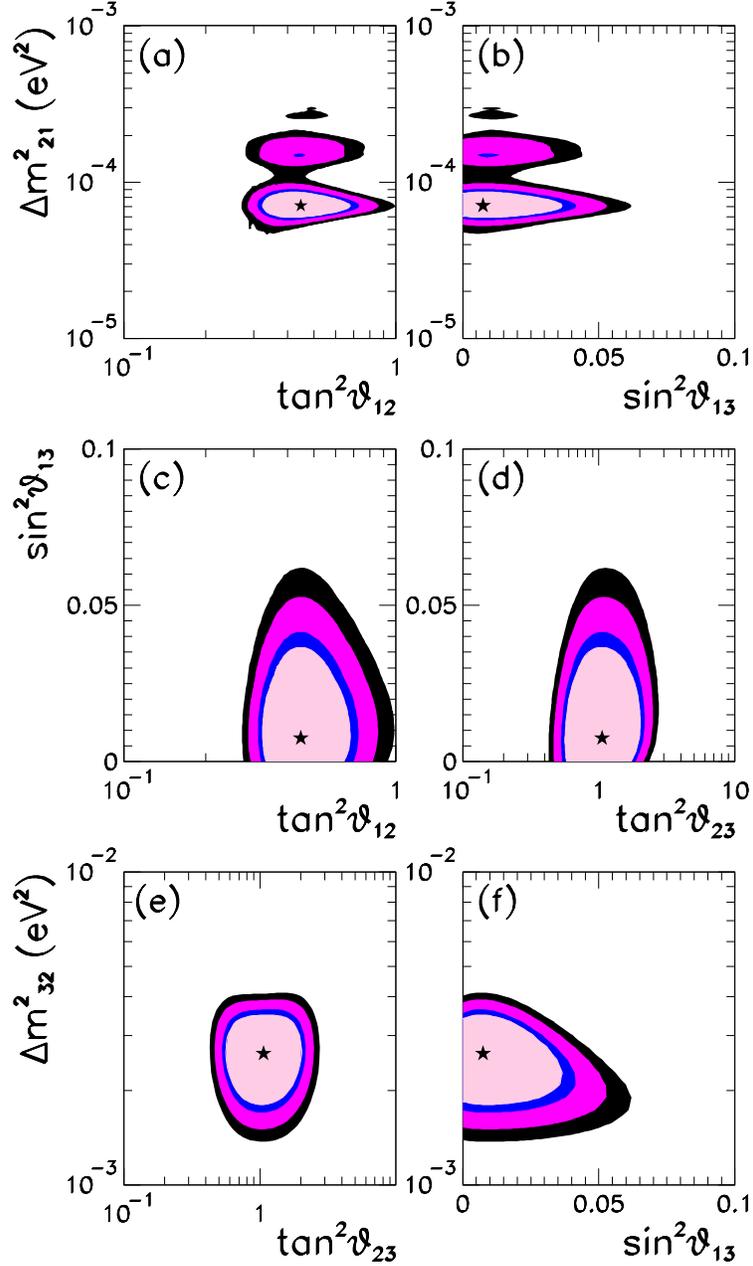,width=4in}} 
\caption{Global 3$\nu$ oscillation analysis. Each panels shows 2-dimensional 
projection of the allowed 5-dimensional region after marginalization 
with respect to the three undisplayed parameters. 
The different contours correspond to the two-dimensional allowed regions 
at 90, 95, 99 \% and 3$\sigma$ CL.}
\label{fig:contours}
\end{figure}

We calculate the global $\chi^2$ by fitting all the available data
\begin{eqnarray}
\chi^2_{\rm global}(\Delta m^2_{21},\Delta m^2_{32},
\theta_{12},\theta_{23},\theta_{13})&=&
\chi^2_{\rm solar}(\Delta m^2_{21},\theta_{12},\theta_{13})+ 
\chi^2_{\rm Kland}(\Delta m^2_{21},\theta_{12},\theta_{13}) 
\nonumber \\
&&
+\chi^2_{\rm atm}(\Delta m^2_{32},\theta_{23},\theta_{13}) +
\chi^2_{\rm K2K}(\Delta m^2_{32},\theta_{23},\theta_{13}) 
\nonumber\\
&& 
+ \chi^2_{\rm CHOOZ} (\Delta m^2_{32},\theta_{13})\; .
\label{eq:chiglobal}
\end{eqnarray}
The results of the global combined analysis are summarized in
Fig.~\ref{fig:chiglo} and Fig.~\ref{fig:contours} in which
we show different projections of the allowed 5-dimensional
parameter space.

In Fig.~\ref{fig:chiglo} we plot the individual bounds on each
of the five parameters derived from the global analysis 
(full line). To illustrate the impact of the K2K and KamLAND
data we also show the corresponding bounds when K2K is not 
included in the analysis (dotted line)  and when KamLAND is
not included (dashed line). 
In each panel the displayed $\chi^2$ has been marginalized with respect
to the other four parameters.  The subtracted minima for each
of the curves are given in Table~\ref{tab:minima}. 

Figure~\ref{fig:chiglo} illustrates that the dominant effects of including
KamLAND  are those derived in the two-neutrino oscillation
analysis of solar and KamLAND data~\cite{ourkland,othersolkam}:
the determination of $\Delta m^2_{21}$ (panel a) to be in the LMA region and
a very mild improvement of the allowed mixing
angle $\theta_{12}$ (panel b). In other words, the inclusion of the 3$\nu$ 
mixing structure in the analysis of solar and KamLAND data does not affect
the determination of these parameters once the additional angle $\theta_{13}$
is bounded to be small. The slight tightening of the $\theta_{13}$ 
limit due   to the inclusion of K2K data does not have any impact in 
the determination 
of the bounds on $\Delta m^2_{21}$ and $\theta_{12}$. Quantitatively we find 
that the following ranges of  parameters are allowed at 1$\sigma$ (3$\sigma$) 
CL from this analysis
\begin{eqnarray}
&(5.4)\; 6.7 \; <\; \Delta m^2_{21}/10^{-5}\mbox{\rm eV$^2$}\;< \; 
\;7.7\, (10.)\;  \;\; {\rm and}\;\;
(14.)\; <\; \Delta m^2_{21}/10^{-5}\mbox{\rm eV$^2$}\;< \; 
(19.)\; , &\nonumber \\
&(0.29)\, 0.39\;<\;\tan^2\theta_{12}\;<\; 0.51 \, (0.82)\; 
\label{skranges}
\end{eqnarray}
The range of  $\Delta m^2_{21}$ on the right of the first
line in Eq.~(\ref{skranges})
correspond to solutions in
the upper LMA island (see Fig.~\ref{fig:contours}.a). At present the 
results of the solar and Kamland analysis still allow for this ambiguity
in the determination of  $\Delta m^2_{21}$  at CL $\gtrsim 2.5\sigma$.
This reflects the departure from the parabolic (gaussian) behaviour
of the $\Delta m^2_{21}$ dependence of  $\chi_{\rm global}$ 
and the presence of a second local minima. With improved statistics
KamLAND will be able to resolve this ambiguity~\cite{othersolkam,carlosnew}.

Comparing the full line on the $\theta_{13}$ panel in 
Fig.~\ref{fig:chiglo}.c with the corresponding one in Fig.~\ref{fig:k2k3fam}.c
we see that the inclusion of the solar+KamLAND data does have an impact
on the allowed range of $\theta_{13}$. However the comparison of 
the full and dashed line in Fig.~\ref{fig:chiglo}.c illustrates that 
the impact is due to the solar data. 
Eq.~(\ref{pee3}) shows that a small $\theta_{13}$ does not 
significantly affect the shape of the measured spectrum 
at KamLAND. On the other hand, the overall normalization is scaled by 
$\cos^4\theta_{13}$ and this factor has the potential to introduce a 
non-negligible effect (in particular in the determination of the mixing angle 
$\theta_{12}$ ~\cite{3kland}). Within its present
accuracy, however, the KamLAND experiment cannot provide any further
significant constraint on $\theta_{13}$.
Altogether, the derived bounds on $\theta_{13}$ from the global analysis are:
\begin{equation}
\sin^2\theta_{13}<0.02 \; (0.052)\; , 
\label{13ranges}
\end{equation}
at 1$\sigma$  (3$\sigma$). 

Finally, comparing Fig.~\ref{fig:chiglo}.d and Fig.~\ref{fig:chiglo}.e with the 
corresponding curves in Fig.~\ref{fig:k2k3fam}.a and Fig.~\ref{fig:k2k3fam}.b,
we see that the additional restriction on the possible range of $\theta_{13} $
imposed by the solar data does not quantitatively affect the 
dominant effect of the inclusion K2K --the improved determination
of $\Delta m^2_{32}$. Thus the allowed ranges of $\Delta m^2_{32}$ 
and  $\theta_{23}$ in Eq.~(\ref{akcranges}) are valid for the global
analysis as well. 

The ranges in Eqs.~(\ref{akcranges}), (\ref{skranges}) and~(\ref{13ranges})
are not independent. In Fig.~\ref{fig:contours} we plot the correlated 
bounds from the global analysis for each pair of parameters. The regions in 
each panel are obtained after marginalization of $\chi^2_{\rm global}$ in
Eq~(\ref{eq:chiglobal}) with respect to the three undisplayed parameters.
The different contours correspond to regions defined at 90, 95, 99 \% 
and 3$\sigma$ CL 
for 2d.o.f ($\Delta\chi^2=4.61,5.99,9.21,11.83$) respectively. From the figure
we see that the stronger correlation appears between $\theta_{13}$ and 
$\Delta m^2_{32}$ as a reflection of the CHOOZ bound.
\begin{table}
\begin{tabular}{|c|c c c|}
\hline
& GLOBAL & GLOBAL-K2K & GLOBAL-KLAND\\
\hline 
Data points & 178 & 172  &  165 \\
$\frac{\Delta m^2_{21}}{\rm eV^2}$ &  $7.1\times 10^{-5}$ & 
  $7.1\times 10^{-5}$ &   $5.8\times 10^{-5}$  \\
$\tan^2\theta_{12}$  &  0.45      &   0.45   & 0.45    \\
$\frac{\Delta m^2_{32}}{\rm eV^2}$ &  $2.6\times 10^{-3}$ & 
$2.5\times 10^{-3}$ & $2.6\times 10^{-3}$   \\ 
$\tan^2\theta_{23}$  & 1      & 1   &   1 \\
$\sin^2\theta_{13}$  & 0.009  & 0.009 &  0.009    \\
$\chi^2_{\rm min}$   & 136  & 127     & 130 \\
\hline
\end{tabular}
\caption{Minimum $\chi^2$ values and best-fit points for
the global $3\nu$ oscillation analysis.}
\label{tab:minima}
\end{table}
In general, because of the correlations, the ranges in 
Eqs.~(\ref{akcranges}), (\ref{skranges}) and~(\ref{13ranges})
cannot be directly used in deriving the corresponding entries in the
$U$ mixing matrix as we discuss next.

\subsection{Determination of the Leptonic Mixing Matrix}
\label{matrix}
We describe in this section our procedure to consistently derive 
the allowed ranges for the magnitude  of the entries of the leptonic mixing 
matrix. We start by defining the mass-marginalized $\chi^2$ function:
\begin{equation}
\chi^2_{\rm mix, global}(\theta_{12},\theta_{23},\theta_{13})=
{\rm min}_{(\Delta m^2_{21},\Delta m^2_{32})} 
\chi^2_{\rm global}(\Delta m^2_{21},\Delta m^2_{32},
\theta_{12},\theta_{23},\theta_{13}) \; .
\end{equation}
We study the variation of $\chi^2_{\rm mix,global}$ as function of
each of the mixing combinations in $U$ as follows. For a given magnitude
$\overline{U_{ij}}$ of the entry $U(i,j)$ we define
$\chi^2(\overline{U_{i,j}})$ as the minimum value of $\chi^2_{\rm mix,
global}(\theta_{12},\theta_{23},\theta_{13})$ with the condition
$|U(i,j)(\theta_{12},\theta_{23},\theta_{13})|=\overline{U_{i,j}}$. In
this procedure the phase $\delta$ is allowed to vary freely between 0
and $\pi$.  The allowed range of the magnitude of the entry $ij$ at a
given CL is then defined as the values $\overline{U_{i,j}}$ verifying
\begin{equation}
\chi^2(\overline{U_{i,j}})-\chi^2_{\rm global,min}\leq
\Delta\chi^2({\rm CL,1d.o.f})\; , 
\end{equation}
with $\chi^2_{\rm global,min}=136$. This is equivalent to having done the 
full analysis in terms of the independent matrix elements -- 
of which, in the hierarchical approximation,  
only three are experimentally accessible at present  
(and can be chosen for instance to be 
$|U_{e2}|$, $|U_{e3}|$, and $|U_{\mu 3}|$)-- 
 and find the allowed magnitude of each
$|U_{ij}|$ by marginalization of  
\begin{equation}
\chi^2_{\rm mix, global}(|U_{e2}|,|U_{e3}|,|U_{\mu 3}|) \; , 
\end{equation}
with the use of unitarity relations and allowing a free relative phase 
$\delta$.

With this procedure we derive the  following 90\% (3$\sigma$) CL
limits on the magnitude of the elements of the complete matrix
\begin{equation}
U=\begin{pmatrix} 
(0.73)\;0.79\;{\rm to}\;0.86\;(0.88)& 
(0.47)\;0.50\;{\rm to}\;0.61\;(0.67)& 
0\;{\rm to}\; 0.16\; (0.23)\cr
(0.17)\;0.24\;{\rm to}\;0.52\;(0.57)&
(0.37)\;0.44\;{\rm to}\;0.69\;(0.73)& 
(0.56)\;0.63\;{\rm to}\;0.79\;(0.84)  \cr
(0.20)\;0.26\;{\rm to}\;0.52\;(0.58)&
(0.40)\;0.47\;{\rm to}\;0.71\;(0.75)& 
(0.54)\;0.60\;{\rm to}\;0.77\;(0.82)
\end{pmatrix}\; .
\label{eq:umatrix}
\end{equation}
By construction the derived limits in Eq.~(\ref{eq:umatrix}) are 
obtained under the assumption of the matrix $U$ being unitary.
In other words, the ranges in the different entries of the matrix
are correlated due to the fact that, in general, the result of a given 
experiment restricts a combination of several entries of the matrix, 
as well as to the constraints imposed by unitarity. 
As a consequence choosing a specific value for one element further 
restricts the range of the others. 

\subsection{$\Delta m^2_{21}$ oscillations at future LBL experiments}
\label{lbl}
In general, correlations between the allowed ranges of the parameters have
to be considered when deriving the present bounds for  any quantity
involving two of more parameters. This is the case, for example,  
when predicting the allowed range of CP violation at future
experiments as discussed in Ref.~\cite{lisi3sol}. 

Here we explore the possible size of 
effects associated with $\Delta m^2_{21}$ oscillations 
(both CP violating and CP conserving) at future LBL
experiments to be performed either with conventional 
superbeams~\cite{SB} 
(conventional meaning from the decay of pions generated 
from a  proton beam dump) 
at a Neutrino Factory~\cite{NF} with neutrino beams from muon decay 
in muon storage rings~. 

The ``golden'' channel at these facilities involve
the observation of either ``wrong-sign'' muons due to
$\nu_e\rightarrow \nu_\mu$ (or $\bar\nu_e\rightarrow \bar\nu_\mu$)
oscillations at a neutrino factory or or the detection of electrons
(positrons) due to $\nu_\mu\rightarrow\nu_e$
($\bar\nu_\mu\rightarrow\bar\nu_e$) at conventional superbeams .  
In either case the relevant
oscillation probabilities in vacuum are accurately given
by~\cite{golden,otherCP}
\begin{eqnarray}
P_{\nu_e\nu_\mu}&=&
s_{23}^2\sin^22\theta_{13}\sin^2
\left(\frac{\Delta m^2_{32}L}{4E}\right)+c_{23}^2 
\sin^22\theta_{12}\left(\frac{\Delta m^2_{21}L}{4E}\right)^2 \nonumber\\ &&
+\tilde J \cos\left(\delta +\frac{\Delta m^2_{32}L}{4E}\right)
\left(\frac{\Delta m^2_{21}L}{4E}\right)
\sin\left(\frac{\Delta m^2_{32}L}{4E}\right)=P^{\rm atm}+P^{\rm sol}+
P^{\rm inter}\; ,   
\label{CP}
\end{eqnarray}
with $\tilde J=c_{13}\sin2\theta_{12}\sin2\theta_{13}\sin2\theta_{23}$.
$P^{\rm sol}$ contains the contribution to the probability
due to longer wavelength oscillations while $P^{\rm inter}$ gives the
interference between the longer and shorter wavelength oscillations and
contains the information on the CP-violating phase $\delta$. In order to
quantify the present bounds on these contributions we factorize the baseline
and energy independent parts as:
\begin{eqnarray}
P^{\rm sol}=(F^{\rm sol})^2\left(\frac{L}{4E}\right)^2 && P^{\rm
inter}=F^{\rm inter} \cos\left(\delta -\frac{\Delta
m^2_{32}L}{4E}\right)\sin\left(\frac{\Delta m^2_{32}L}{4E}\right)
\left(\frac{L}{4E}\right)\; , \nonumber\\ F^{\rm
sol}=c_{23}\sin2\theta_{12}\Delta m^2_{21} && F^{\rm
inter}=c_{13}\sin2\theta_{12}\sin2\theta_{13}\sin2\theta_{23}\Delta
m^2_{21} \;.
\label{coeff}
\end{eqnarray}
For very long baselines, for which  the presence of matter cannot 
be neglected, the expressions above for $F^{\rm sol}$ and 
$F^{\rm inter}$ still hold as the coefficients of the dominant contributions 
to the probabilities in the expansion in the small 
parameters $\theta_{13}$ and $\Delta m^2_{21}$~\cite{golden,otherCP}. 

\begin{figure}[!t]
\centerline{\psfig{figure=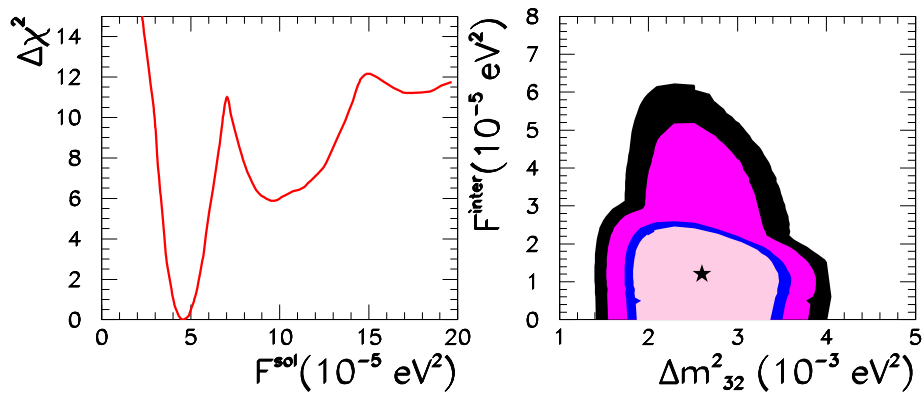,width=5in}} 
\caption{(a) Dependence of $\Delta\chi^2$ on $F^{\rm sol}$. 
(b) Allowed regions 
of $F^{\rm inter}$ versus $\Delta m^2_{32}$ at 90, 95, 99\% and 3$\sigma$.
The best value is marked with a star.} 
\label{fig:lbl}
\end{figure}

We show in Fig.~\ref{fig:lbl} the present bounds on the coefficients 
$F^{\rm sol}$ and $F^{\rm inter}$. 
In general the 
dependence on $\Delta m^2_{32}$ of the interference term cannot be factorized 
because,  depending on the considered baseline and energy,  the oscillating  
phase with $\Delta m^2_{32}$ may be not small enough to be expanded. For this
reason we show in  Fig.~\ref{fig:lbl}.b
the 2-dimensional allowed region  of $F^{inter}$ versus $\Delta m^2_{32}$.
In the figure we mark with a star the best value for  $F^{inter}$ as
obtained from this analysis which is not vanishing due to the 
small but non-zero best fit value of $\sin\theta_{13}$.  
This is, however,  not statistically significant as  $F^{inter}=0$ is at 
$\Delta\chi^2=0.9$. The negative slope in the upper part of the 
90\% and 95\% CL region in  
Fig.~\ref{fig:lbl}.b is a reflection of the anti-correlation between the 
$\Delta m^2_{32}$ and $\sin^2\theta_{13}$ constraints from the CHOOZ 
experiment  (see Fig.~\ref{fig:contours}.f)

From this study we find the following 1$\sigma$ (3$\sigma)$ [1d.o.f] bounds 
\begin{eqnarray}
&F^{\rm sol}/(10^{-5}{\rm eV}^2)&=4.6\pm0.6 \; (4.6^{+2.1}_{-1.6}
\;\;{\rm and}\; 9.5^{+3.5}_{-2.0})\; ,  \label{eq:lblsol}\\
0<&F^{\rm inter}/(10^{-5}{\rm eV}^2)&<1.9 \;(5.5)\; ,  \label{eq:lblint}
\end{eqnarray}
where the bounds on  $F^{\rm inter}$ are shown for the best fit value of 
$\Delta m^2_{32}=0.0026$ eV$^2$. 
The larger values for the 3$\sigma$ range in Eq.(\ref{eq:lblsol}) 
and $F^{\rm inter}$ correspond to solutions of the solar+KamLAND analysis
lying in the higher $\Delta m^2_{21}$ island (see Fig.~\ref{fig:contours}.a 
and discussion below Eq.~(\ref{skranges})).

\section{Summary and Conclusions}

\label{conclu}
We have presented the results of  an updated global analysis  of
solar, atmospheric, reactor and long-baseline neutrino  
data in the context of three-neutrino oscillations
making special emphasis on the impact of the recent 
long baseline $\nu_\mu$  disappearance data from the K2K experiment 
and reactor $\bar\nu_e$ disappearance from the KamLAND experiment.
We find that the dominant effect of the inclusion  
of K2K and KamLAND data is the reduction of the allowed
range of $\Delta m^2_{32}$ and $\Delta m^2_{21}$ respectively while
the impact on the mixing angles $\theta_{23}$ and $\theta_{12}$ is
marginal. The increase of the lower
bound on $\Delta m^2_{32}$ due to the inclusion of the K2K data 
leads also to a slight  tightening of the derived limit on 
$\theta_{13}$ at high CL. 
Our results on the individual allowed ranges for the oscillation parameters
are given in Eqs.~(\ref{akcranges}), (\ref{skranges}) and~(\ref{13ranges})
and graphically displayed in Fig.~\ref{fig:chiglo}. The correlations
between the derived bounds are illustrated i  
in Fig.~\ref{fig:contours}. As outcome of the analysis we have presented
in Eq.~(\ref{eq:umatrix}) our up-to-date best determination of 
the magnitude of the elements of the complete leptonic mixing matrix.
Finally  we have quantified 
the allowed contribution of $\Delta m^2_{21}$
oscillations to CP-odd and CP-even observables at future long 
baseline experiments with results presented 
in Fig.~\ref{fig:lbl} and Eqs.~(\ref{eq:lblsol}) and~(\ref{eq:lblint}).

\section*{APPENDIX A: ANALYSIS OF K2K DATA}
In this appendix we describe our calculation of the K2K spectrum and
our statistical analysis of the K2K data \cite{k2kprl}. 

We use in our statistical analysis the K2K data on the spectrum 
of single-ring $\mu$-like events. K2K present their results as 
the number of observed events as a function of the reconstructed
neutrino energy. The reconstructed neutrino energy  is determined 
from the observed $\mu$ energy in the event, 
$E_\mu$,   and  its scattering angle with respect to the incoming beam 
direction, $\cos\theta_\mu$,  as
\begin{equation}
E_{\rm rec}=
\frac{m_N E_\mu-\frac{m_\mu^2}{2}}{m_N- E_\mu+p_\mu\cos\theta_\mu}\; , 
\end{equation}
where $m_N$ is the nucleon mass.  In Fig.~\ref{fig:k2kdist} we show their
data binned in five 0.5GeV bins with $0<E_{\rm rec}<2.5$ plus 
one bin containing all events above $2.5$ GeV. The total sample 
corresponds to 29 events. In the absence of oscillations 44 events
were expected. 

For QE events, $\nu_\mu\, n\rightarrow \mu \, p$ , and assuming perfect
$E_\mu$ and $\cos\theta_\mu$ determination,  $E_{\rm rec}=E_\nu$. 
Experimental energy and angular resolution, nuclear effects, 
and, more importantly,
the contamination from non-QE (nQE) events, $\nu_\mu N \rightarrow \mu X$, 
in the sample, result into important deviations of the defined 
$E_{\rm rec}$ from the {\sl real} $E_{\nu}$. From simple kinematics
one finds that in nQE events  there is a shift in the
reconstructed neutrino energy with respect to the true neutrino energy
\begin{equation}
E_{\rm rec}= E_\nu 
\left[1 +\frac{M_X^2-m_N^2}{m_N E_\mu-\frac{m_\mu^2}{2}}\right]^{-1} <E_\nu\; , 
\label{eq:erecnqe}
\end{equation}
where $M_X$ is the invariant mass of the hadronic system produced together
with the muon in the $\nu_\mu$ interaction. 
At the K2K energies the most important nQE contamination comes 
from single pion production.
which occurs via the $\Delta$ resonance. At the largest energies 
there is a small contribution from deep inelastic scattering.  

Thus in general the  observed spectrum of single-ring $\mu$-like events
in K2K can be obtained as 
\cite{kobayashi}
\begin{eqnarray}
N^{\rm th} (E_{\rm rec})&=& N_{\rm norm}
\int \Phi_{SK} (E_\nu) \,P_{\mu\mu}(E_\nu) \big[\sigma_{QE}(E_\nu) \,
\epsilon_{QE}^{1R_\mu}(E_\nu) \,r_{QE}(E_\nu, E_{\rm rec}) + \nonumber\\
&& f_{nQE} \, \sigma_{nQE}(E_\nu)\,\epsilon_{nQE}^{1R_\mu}(E_\nu) \,
r_{nQE}(E_\nu, E_{\rm rec}) \big] 
dE_\nu \; , 
\label{eq:dist}
\end{eqnarray}
where $\Phi_{SK}(E_\nu)$ is the expected $\nu_\mu$ spectrum at 
the SK site in the absence of oscillations. 
$ P_{\mu\mu}(E_\nu)$ is the survival probability of $\nu_\mu$
for a given set of oscillation parameters. 
$f_{nQE}=0.93$ is a rescale factor of the expected contamination 
from nQE events as obtained from MC simulation by the K2K 
collaboration \cite{k2kprl}. $\sigma_{QE(nQE)}(E_\nu)$ are the
neutrino interaction cross sections. 
$\epsilon_{QE(nQE)}^{1R_\mu}$ are the detection efficiencies for 1-ring
$\mu$-like events at SK.  
$r_{QE(nQE)}(E_\nu, E_{\rm rec})$ 
are the functions relating the reconstructed
energy and the true neutrino energy. $N_{\rm norm}$ is the
normalization factor which is chosen so that in the absence 
of oscillations the total integral gives 44 events. 

In our calculation we use
the neutrino spectrum  $\Phi_{SK}$ as provided by the K2K collaboration
\cite{kobayashi,kobaprivate}. This flux was estimated from the flux
measured in the near detector by multiplying it by a MC simulated
ratio of the fluxes between the near and far detectors. 
We further assume that the detection efficiencies for 
1-ring $\mu$-like events at SK are the same for the K2K
analysis as for the atmospheric neutrino analysis
(further details and references can be found in 
Ref.~\cite{ouratmos}). 

At present there is not enough information from the K2K 
collaboration on the 
$r_{QE(nQE)}(E_\nu, E_{\rm rec})$  functions. In our calculation we have used
a  {\sl physically-motivated} form for those functions. 
We include in 
the functions $r_{QE(nQE)}(E_\nu, E_{\rm rec})$ the dominant effect 
in the missreconstruction of the 
neutrino energy: the shift in the reconstructed neutrino energy 
due to the different kinematics of the nQE events as described by
Eq.~(\ref{eq:erecnqe}). We also include the
(subdominant) effects due to the experimental energy and angular 
resolutions which smear the measured muon energy $E_\mu$ and angle 
$\theta_\mu$ around their true values $E'_\mu$  and $\theta'_\mu$:
\begin{eqnarray}
r_{QE}(E_\nu, E_{\rm rec}) =
\frac{1}{\sigma_{QE}(E_\nu)}
&\int& dE'_\mu  dE_\mu d\theta 
\frac{d\sigma_{QE}(E_\nu, E'_\mu) }{dE'_\mu}
{\rm Res_E}(E_\mu-E_\mu'){\rm Res_\theta}(\theta_\mu-\theta_\mu')
\delta(E_\nu-E'_{\rm rec})
\nonumber \\
r_{nQE}(E_\nu, E_{\rm rec}) =
\frac{1}{\sigma_{nQE}(E_\nu)}
&\int&  dE'_\mu  dE_\mu d\theta 
dM_X \frac{d\sigma_{nQE}(E_\nu, E'_\mu,M_X) }{dE'_\mu dM_X}
{\rm Res_E}(E_\mu-E_\mu'){\rm Res_\theta}(\theta_\mu-\theta_\mu') 
\nonumber \\
&& 
\delta \left(E_\nu-E'_{\rm  rec} 
\left[1 +\frac{M_X^2-m_N^2}{m_N E'_\mu-\frac{m_\mu^2}{2}}\right]^{-1}
\right) 
\label{eq:reso1}
\end{eqnarray}
where 
\begin{eqnarray} 
E'_{\rm  rec}= E_{\rm rec} 
\frac{m_N E_\mu-\frac{m_\mu^2}{2}}{m_N-E_\mu+p_\mu\cos\theta'_\mu}
\frac{m_N -E'_\mu+p'_\mu\cos\theta_\mu}{m_N E'_\mu-\frac{m_\mu^2}{2}} \; , 
\nonumber \\
{\rm Res_E}(E'_\nu-E_\nu)=\frac{1}{\sqrt{2\pi}\sigma_E} 
{\mbox{\Large e}}^{-\frac{1}{2}\frac{(E_\mu-E_\mu')^2}{\sigma_E^2} }\; , 
\nonumber \\
{\rm Res_\theta}(\theta'_\nu-\theta_\nu)=\frac{1}{\sqrt{2\pi}\sigma_\theta} 
{\mbox{\Large e}}^{-\frac{1}{2}\frac{(\theta_\mu-\theta_\mu')^2}
{\sigma_\theta^2}}\; . 
\label{eq:reso2}
\end{eqnarray}
Following SK data \cite{skatmpub} we use an energy resolution for the
muons of $\sigma_E/E_\mu=3$\% and an angular resolution 
$\sigma_\theta=3^\circ$ (see also Ref.~\cite{campanelli}
for further details). Notice that in the expressions above 
the true angle of the muon,
$\theta_\mu'$, is not an independent variable but it is related by
the kinematics of the process  to the  initial neutrino energy, $E_\nu$,
the final muon energy,  $E_\mu'$, and the invariant mass of the 
hadronic system, $M_X$.
The final result on the number of expected events in each $E_{\rm res}$
bin is obtained by substituting Eqs.~(\ref{eq:reso1}) and~(\ref{eq:reso2}) in 
Eq.~(\ref{eq:dist})  and numerically integrating for the 
kinematical variables in the corresponding range of $E_{\rm res}$. In 
this procedure the only free parameter to
adjust is the overall normalization. The shape of the spectrum is
then fully determined.

In order to verify the quality of our simulation we compare our
predictions for the energy distribution of the events with the Monte
Carlo simulations of the K2K collaboration in absence of
oscillation. In Fig.~\ref{fig:k2kdist} 
we show our predictions superimposed with those from the experimental 
Monte Carlo (obtained from Fig.2 in Ref.~\cite{k2kprl}), both normalized 
to the 44  expected events in the absence of oscillations. 
The boxes  for the MC prediction represent the systematic error bands. 
We can see that the agreement in the shape of the spectrum is very good. 

In our statistical analysis of the K2K data we use  Poisson statistics
as required given the small number of events. 
We include the systematic uncertainties associated with the determination 
of the neutrino energy  spectrum in the near detector (ND), 
the model dependence of the amount of nQE contamination parameter $f_{nQE}$, 
the near/far extrapolation (F/N) 
and the overall flux normalization (NOR) \cite{kobayashi,kobaprivate}. 
The errors on the first three items depend on energy and have correlations
among the different energy bins.
We account for all these effects by using the $\chi^2$ 
function~\cite{kobayashi,lisik2k,kobaprivate}:
\begin{eqnarray}
\chi^2_{K2K}={\rm min}_{f} 
&\Big[& 2\sum_{i=1}^6\left(\bar
N_i^{\rm th}-N_i^{\rm exp} - N_i^{\rm exp}\ln
\frac{\bar N_i^{\rm th}}{N_i^{\rm exp}}\right)+
\sum_{j,k=1}^{6} f^{F/N}_j (\rho^{F/N})^{-1}_{ij} f^{F/N}_j 
\nonumber \\
&&+\sum_{j,k=1}^{7} f^{ND,nQE}_j (\rho^{ND,nQE})^{-1}_{ij} f^{ND,nQE}_j 
+{f^{NOR}}^2 \Big] \; , 
\end{eqnarray}
where $ \bar N_i^{\rm th}= N_i^{\rm th} (1+ f^{F/N}_i \sigma^{F/N}_i +
f^{ND}_i \sigma^{ND}_i + f^{nQE} \sigma^{nQE}_i+f^{NOR} \sigma^{NOR})$.
By ${\rm min}_{f}$ we denote the minimization with respect to the
systematic shift parameters (or pulls~\cite{lisik2k}) 
$f^{F/N}_{i=1,6}$, $f^{ND}_{i=1,6}$, $f^{nQE}$, and $f^{NOR}$
We use the systematic errors and their correlations as provided by 
K2K collaboration \cite{k2kprl,kobayashi,kobaprivate}. For instance   
\begin{eqnarray}
\sigma^{NOR}&=&5\% \; , \nonumber\\
\sigma^{F/N}_i&=&
2.5\%\;,\; 4.3\%\;,\; 6.5\% \; ,\; 10.4 \%\,\; 11.1\%\;, 12.2 \%\;,  
\nonumber\\   
\sigma^{ND}_i &=&
49\%\;,\; 7.1\%\;,\; 0\% \; ,\; 7.1 \%\,\; 8.4\%\;, 11.1 \%\;,  \\
\sigma^{nQE}_i&=& 13\%\;,\; 8.9\%\;,\; 6\% \; ,\; 3.8 \%\,\; 3.\%\;, 5.5 
\nonumber
\end{eqnarray}
for $i=1,6$ respectively. 

Thus in our analysis  we use both the shape and the normalization of 
the 29 single-ring $\mu$-like events.  
In their analysis  K2K uses only the spectrum shape 
(but not the normalization) of the 29 single-ring $\mu$-like events plus 
the overall normalization 
of their total sample of fully contained events (a total of 56). 
We cannot use the normalization from the additional 27 events in the lack of 
more detailed information from the
K2K collaboration on the efficiencies for multi-ring events.  
Nevertheless, as described in Sec.~\ref{k2k}, the results of our oscillation 
analysis are in good agreement with those from the K2K analysis.

\acknowledgments 
We are particularly indebted to T. Kobayashi for providing us with 
information and clarifications about the K2K analysis. 
We thank M. Maltoni for many useful discussions and for his collaboration
on the atmospheric neutrino analysis during the early stages of this work.
This work was supported in part by the National Science Foundation grant 
PHY0098527. CPG acknowledges support from NSF grant No. PHY0070928. 
MCG-G is also supported  by Spanish Grants No FPA-2001-3031 
and CTIDIB/2002/24.

\end{document}